\documentclass[aps,prb,twocolumn]{revtex4}
\usepackage{epsfig}
\usepackage{amsmath}
\begin{document}

\title{On the Existence of Prewetting in Supracritical Fluid Mixtures,}
\author{ {Jan Forsman* and Clifford E. Woodward**}
\\
{*}Theoretical Chemistry, Chemical Centre
\\
 P.O.Box 124, S-221 00 Lund, Sweden
\\
{**}School of Science
\\
University of NSW, Canberra
\\
ACT 2600, Australia   
\\
}
\begin{abstract}
In this communication we demonstrate the existence of a first-order prewetting transition 
of a supracritical model polymer solution adjacent to an attractive surface. The model fluid we use
mimics (qualitatively) an aqueous polyethylene oxide solution and, like the actual solution, 
displays a closed loop 2-phase region with an upper and lower critical solution temperature.
The model fluid is shown to undergo a prewetting transition at an adjacent attractive surface
even at temperatures below the lower critical solution temperature (supracriticality).  This phenomenon
follows from non-local thermodynamics when the lengthscale of the relevant fluid 
structures of surface films are commensurate or smaller than the range of intermolecular interactions. 
\end{abstract}
\maketitle
\newpage

Many binary fluid mixtures or solutions display a demixing phase diagram 
with both an {\em upper}  and a {\em lower}  critical solution temperature ({\em U/LCST}),
which bracket the 2-phase region.   Above the {\em UCST} and below the {\em LCST}, in what we will
refer to as the {\em supracritical} regions,
the mixture exists as a single phase.  In this article, we will be primarily concerned with the supracritical  
region below the {\em LCST}.  

The presence of an {\em LCST} is not common in mixtures of simple fluids, but in polymer solutions it is more the rule rather than the exception.
\cite{Kleintjens:80,vanKoynenburg:80,Nakanishi:82,Evans:86,Burgess:89,Karlstrom:85,Jackson:91,Morshige:97,Dormidontova:02,Drzewinski:09}.
One example is an aqueous solution of poly(ethylene oxide) (PEO)
which has an {\em LCST} that approaches 100$^\circ$C at high molecular weights \cite{Saeki:73}.
Using  {\em classical} polymer density functional theory (cDFT), we have generalised a successful
theoretical model for such solutions \cite{Karlstrom:85}, that can be used to model such solutions in
heterogeneous environments\cite{Xie:13,Xie:16a,Haddadi:21b}. 
In this model, monomers are assumed to be in either of two classes of states, labelled {\em A} and {\em B}, 
where {\em B} is more solvophobic than {\em A}. Thus A states become preferred as the temperature is lowered
which promotes complete dissolution.  On the other hand, the degeneracy of the {\em B} class exceeds that of {\em A}, 
which means the relative probabilty of B (solvophobic) monomers {\em increases} with temperature.  This
creates the possibility of  demixing with increasing temperature, leading to an {\em LCST}.  At a high enough temperature, 
the mixing entropy dominates interactions, giving rise to an {\em UCST}.   We do not
resort to temperature-dependent interactions {\em per se}, but instead make explicit the intrinsic enthalpic and
entropic contributions of {\em A} and {\em B} species within an aqueous environment.

Consider a bulk mixture which is able to diffuse 
freely into and out of a pore.  If the pore surfaces interact
directly with the confined fluid, they may cause a phase change relative to
the bulk fluid, so-called capillary induced phase separation (CIPS) \cite{Wennerstrom:98}.  
If we make the assumption that the fluid-fluid interactions are 
short-ranged compared to the size of the pore then the pore 
will only contribute surface terms to the free energy and
CIPS can only occur  if the bulk fluid is {\em not} supracritical.   
In pores which are narrow compared to the range of fluid-fluid interactions, there may be a
significant shift along the temperature axis of the 2-phase region of the confined fluid.
This is because surfaces will truncate intermolecular interactions between
fluid particles\cite{Lane:81,Burgess:89,Wennerstrom:98,Kotelyanskii:98}.
As these interactions are generally attractive at long range (at least for non-ionic fluids)
this causes a decrease in the cohesive forces within the fluid.  
To illustrate, consider a single-component fluid displaying a critical temperature.
Reduction in the cohesive forces, due to truncation, will mean 
the econfined fluid behaves similarly to that of the bulk 
fluid, but at a higher temperature. Thus the overall phase diagram of the
confined fluid is shifted to lower temperatures, compared to that of the bulk \cite{Burgess:89}.
Such a situation is still consistent with the requirement above, i.e., CIPS can only occur if the 
temperature of the bulk fluid is below the critical temperature.  

However, the scenario is more complex for mixtures that display an {\em LCST} in the bulk.
In previous work, we have shown that confining the mixture in a pore 
reduces the {\em LCST} \cite{Xie:13,Xie:16a,Haddadi:21b}. Thus, in this case,  
the confined fluid may undergo CIPS, even though the bulk solution is {\em supra-critical}. 
Such transitions may occur with inert surfaces, which are repulsive to the solution components \cite{Haddadi:21b}.

In the presence of a single surface, it would seem that the the semi-infinite geometry would only affect 
surface contributions to the free energy of the adjacent fluid and there will be no shift in the 2-phase region.   
However, some fluids may undergo first-order phase transitions at a single surface
via so-called ``thin-thick'' (prewetting) transitions.  
These are {\em surface} transitions, as the  prewetting phases 
involve structural changes of the fluid that are intrinsically confined to narrow films adjacent 
to the adsorbing surface.  Thus, the truncation of fluid-fluid interactions discussed above will plausibly 
also affect prewetting transitions.  In particular, according to the discussion above,
one expects that the {\em UCST} of the prewetting transition is lower than that of the
bulk phase transition.  On the other hand, the role of these mechanisms in the vicinity
of an {\em LCST} has hitherto not attracted  significant attention in the literature, as far as we are aware.   

Consider a bulk 2-component solution, at some fixed pressure and temperature, $T$.
This mixture possesses a demixing regime wherein the fluid 
separates into a concentrated, {\em C}, and dilute, {\em D}, phase
with respect to one of the components (the "solute" species).  For the moment,
we will assume that this solution displays only a {\em UCST}. 
Suppose such a solution in its undersaturated {\em D} phase, 
$T < ${\em UCST}, is adjacent to a single surface, {\em W}, which is attractive 
to the solute species. A positive adsorption of solute at the surface
gives rise to what we will label a "thin" surface film of excess solute.
The surface tension at the surface-fluid interface is denoted as,  $\gamma_{WD}$.  
Making the bulk fluid more concentrated will cause an increase in the excess solute adsorption.
Above the so-called wetting temperature, $T_W$, this increasing adsorption 
will grow by way of a thickening film, with width $L$.  The fluid in this film
will have a character similar to the metastable $C$ phase and the surface-fluid interfacial tension 
will approach  $\gamma_{WC}$.  This occurs because $\gamma_{WC} < \gamma_{WD}$ and 
the free energy is lowered accordingly.   However, the film also establishes a $C-D$ interface with the
bulk and the ensuing positive surface tension contribution, $\gamma_{CD}$, will act to counter 
the free energy lowering at the adsorbing surface. 
At saturation (i.e., on the bulk coexistence curve), the free energy penalty 
for growing a film ($L \rightarrow \infty$) approaches $\Delta \gamma$ (= $\gamma_{WC} + \gamma_{CD} -\gamma_{WD}$),
from above, where  the surface tensions are defined with respect to the coexisting $C$ and $D$ phases. 
For $T > T_W$,  $\gamma_{CD}$ is sufficiently small so that, $\Delta \gamma<0$, 
and the film grows to infinity spontaneously.  This gives rise to a divergence in the excess adsorption of solute
and {\em complete} wetting.  For $T <  T_W$, $\gamma_{CD}$ is large  enough that 
$\Delta \gamma>0$: the thin film persists and no wetting layer forms (partial wetting).  
 
In the above we suggested film growth is continuous as the bulk approaches saturation.
In some cases, the film thickness may grow discontinuously, undergoing a 
first-order, so-called "thin-thick" (or prewetting) transition, at some undersaturated value of {\em D}.  
The locus of concentrations where this transition occurs with varying $T$, is called the prewetting 
line.  This line intercepts the bulk coexistence line at $T_W$, and terminates at some upper {\em critical 
prewetting temperature} {\em UCST}$_{pw} < ${\em UCST}.  This last inequality occurs due to the
truncation mechanism described earlier, which applies here because of the finite width of the 
fluid films that coexist on the prewetting line.  
The free energy (per unit area) cost for film formation at $T > T_W$ in 
an {\em undersaturated} bulk solution can be written as 
$\Delta \gamma+\Delta F_{CD}(L)$.  Here, $\Delta F_{CD}(L)>0$
is essentially the intrinsic free energy of the thick fluid film, without the
contributions from the W-fluid and fluid-fluid surface terms.  If 
$\Delta f_{CD}$ denotes the difference in free energy per unit volume of 
the metastable bulk $C$ phase relative to the stable bulk $D$ phase 
(at the same temperature and pressure), then we have $\Delta F_{CD}(L)\approx \Delta f_{CD}L$
for large enough $L$.  The prewetting transition occurs because 
the system admits two stable film phases on the prewetting line.  For $T = T_W$ 
we have $\Delta F_{CD}(L) = \Delta \gamma = 0$. 

\begin{figure}[h!]
\begin{center}
       \includegraphics[width=7.8cm]{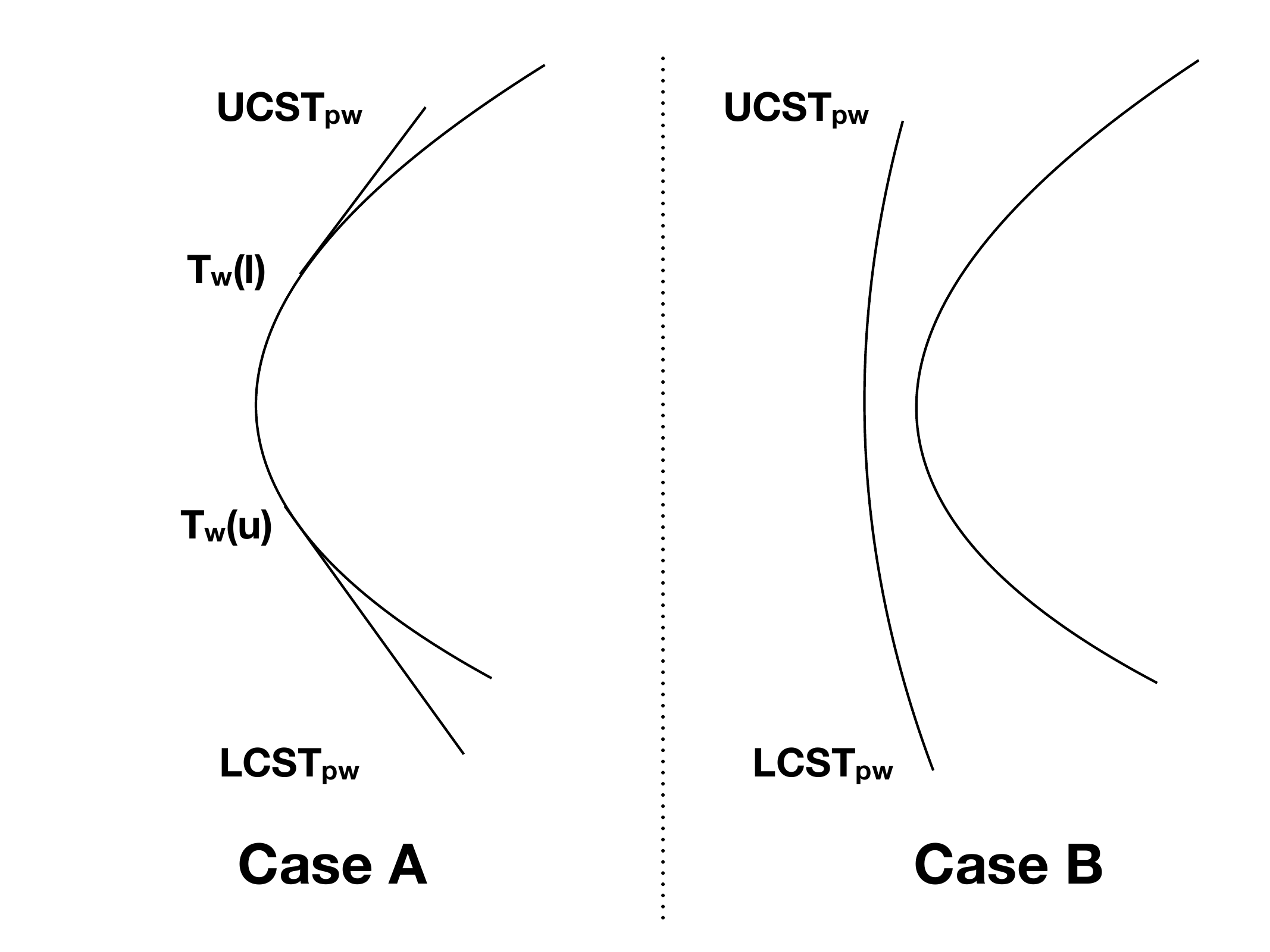}
       \caption{\footnotesize{An illustration of two possible scenarios. In ``Case A'', there are two
           prewetting lines, that terminate at the bulk coexistence line: one with an
           upper wetting temperature, $T_w(u)$, and the other
           with a lower wetting temperature, $T_w(l)$. In ``Case B'', the entire prewetting line
           is shifted from bulk coexistence (undersaturated), and is bound by
           an upper ({\em UCST}$_{pw}$) and lower ({\em LCST}$_{pw}$) critical prewetting temperature.
           The former is always below the bulk {\em UCST}, but {\em LCST}$_{pw}$ may be located below the
           bulk {\em LCST}.
}}
\label{fig:cartoon}
\end{center}
\end{figure}

So what happens for a mixture which also possesses an {\em LCST} as well? 
As argued above for the case of a {\em UCST} only, $T_W$, primarily arises due to the increase in $\gamma_{CD}$
as the temperature decreases.  This is correlated with the increasing compositional differences between coexisting 
bulk $C$ and $D$ phases, as the temperature decreases.  When an {\em LCST}
is present, the 2-phase region becomes a closed loop, which causes $\gamma_{CD}$ 
to decrease again as $T$ decreases toward the {\em LCST}.  This gives rise to the
possibility that both an upper and lower wetting temperature exist, giving rise to two prewetting
lines, respectively terminating above at {\em UCST}$_{pw}$ and below at {\em LCST}$_{pw}$. 
 On the other hand, it is possible that the two prewetting lines will detach from the coexistence 
 line and the system has a single prewetting line of finite length and no wetting temperature, as shown 
 in Figure \ref{fig:cartoon}.
 Furthermore, due to truncation of intermolecular forces in the finite films of prewetting phases, 
as described above, we also expect that  {\em LCST}$_{pw} < ${\em LCST}, see Figure \ref{fig:cartoon}.  This suggests the 
fascinating possibility of thin-thick transitions under supracritical conditions ({\em supracritical prewetting}). That is, 
at some temperatures, lower than {\em LCST}, prewetting transitions occur but don't signal the onset of 
film growth on the surface as the bulk solution concentration increases.  Thus the excess 
solute adsorption always remains finite at all bulk concentrations. In this sense, the term "prewetting transition"
becomes something of a misnomer.  That is, while the system can undergo a thin-thick transition it does not
lead to complete wetting as the composition increases.  This underlines the fact that the prewetting transition 
is essentially a surface transition and is not intrinsically dependent on complete wetting in any fundamental sense.   

Following this qualitative discussion, we will now turn to quantitative predictions.  In particular, we will use the
aqueous PEO solution model described earlier.  This will 
require some model details, and definitions. We let $N({\bf R})$ denote the density distribution
of monodisperse PEO polymers with $r$ monomers.  Here,
${\bf R} = {\bf r}_1,{\bf r}_2,...,{\bf r}_r$, where ${\bf r}_i$ is the coordinate 
of monomer $i$. The polymers are assumed linear with monomers joined
by freely rotating bonds, described by a potential $V_b$. This potential is chosen to ensure a constant
bond length $b$, i.e. $e^{-\beta V_b({\bf R})} \propto \prod \delta(|{\bf r}_{i+1}-{\bf r}_i|-b)$,
where  $\delta (x)$ is the Dirac delta function, and $\beta = 1/(kT)$ is the inverse thermal energy.
We will assume that the total mixture (polymer and solvent) is incompressible. This is ensured by constraining
the total density of monomers+solvent  particles to a fixed value, $n_t$, i.e.
$n_t = n({\bf r})+n_S({\bf r})$, where $n({\bf r})$ and $n_S({\bf r})$ are the monomer and solvent densities at position ${\bf r}$.
Specifically, we have set $n_t\sigma^3 = 1$, where $\sigma$ is
the Lennard-Jones length parameter, common to all species, as defined below.
According to our description above, we introduce the monomer {\em state} probabilities $P_A$ and $P_B$, and defining
state densities $n_A({\bf r}) \equiv n({\bf r}) P_A({\bf r})$ and $n_B({\bf r}) \equiv n({\bf r}) (1-P_A({\bf r}))$,
we write the fluid free energy density functional,  $F$, as:
\begin{eqnarray}
\beta F & = & \int N({\bf R}) \left(\ln [N({\bf R})] - 1 \right) d{\bf R} + \beta \int N({\bf R}) V_b({\bf R}) d{\bf R} + \nonumber \\
               &   &\int \left(n_S({\bf r})\ln\left[n_S({\bf r})\right] + n({\bf r})\right) d{\bf r} + \nonumber \\
&   &\frac{\beta}{2}\sum_{\alpha,\beta}\int\int n_\alpha({\bf r}) n_\beta({\bf r'}) \phi^{(a)}_{\alpha\beta}(|{\bf r}-{\bf r'})| d{\bf r'}d{\bf r}+ \nonumber \\
               &   &\int n_A({\bf r})\ln\left[\frac{P_A({\bf r})}{g_A}\right]d{\bf r} + \nonumber \\
               &   &\int n_B({\bf r})\ln\left[\frac{1-P_A({\bf r})}{g_B}\right]d{\bf r} + \nonumber \\                    
               &   &\beta  \sum_{\alpha,\beta}\int n_\alpha({\bf r})V_{ex}^{\alpha}({\bf r}) d{\bf r}
\label{eq:freen}
\end{eqnarray}
where we have included an external (surface) potential, $V_{ex}^{\alpha}({\bf r})$, 
that acts on the $\alpha$ particles, with $\alpha = A$, $B$ or $S$.  

All particles (monomers and solvent) particles have a hard core with diameter, $\sigma$,
and interact with each other via a Lennard-Jones (L-J) potential, $\phi_{\alpha\beta}^{(a)}(r)$:
\begin{equation}
\phi^{(a)}_{\alpha\beta}(r) = 4 \epsilon_{\alpha\beta}((\frac{\sigma}{r})^{12}-(\frac{\sigma}{r})^6), \hspace{36pt} r > \sigma
\end{equation}
Defining a reference energy parameter, $\epsilon_{ref}$, we obtain reduced
parameters $\epsilon_{\alpha\beta}^* \equiv \epsilon_{\alpha\beta}/\epsilon_{ref}$, 
as well as a reduced temperature $T^* = kT/\epsilon_{ref}$.  We will use
the same energy parameters as in our previous studies.  \cite{Xie:13,Xie:16a,Haddadi:21b}
In that model, the parameters are shifted so that
all long-range $AA$, $AS$ and $SS$ interactions vanish, i.e.,
$\epsilon_{AA}^*  = \epsilon_{SS}^* = \epsilon_{AS}^* =  0$, while
$\epsilon_{BB}^*  = \epsilon_{SB}^*  =  -0.7$, and $\epsilon_{AB}^* = \sqrt{0.3}-1 \approx -0.45$.
We have previously shown that for $g_A/g_B = 13$, one obtains a bulk fluid
with both an {\em UCST} and a {\em LCST}, for long enough chains \cite{Xie:16a}.

The potential function, $V_{ex}^{\alpha}({\bf r})$, defines the nature of the surface, which is
modelled as a hard flat wall parallel to the $(x,y)$ plane (with $z$ normal).  We set:
\begin{equation}
V_{ex}^{\alpha}(z)= \left[
\begin{array} {ll}
\infty, & z < 0 \\
W_\alpha\left[w(z)-w(z_c)\right], & 0 < z < z_c \\
0, & z > z_c \\
\end{array} \right.
\label{eq:wallpot}
\end{equation}
where $w(z) = (1-e^{-z/\sigma})^2-1$, and $z_c = 10\sigma$.
In this case, we chose $W_A = W_S = 0$, which allows us to regulate
the surface properties by a single parameter, $W_B^* \equiv W_B/\epsilon_{ref}$, that
determines the affinity to {\em B}.

The grand potential is defined as, $\Omega = F/A - \int\mu_p n(z) dz$,
where $\mu_m$ is the monomer chemical potential and $A$ is the surface area
of the walls, which we assume is infinite. 
The functional,  $\Omega$, is then minimised with 
respect to both $n(z)$ and $P_{\alpha}(z)$ to obtain the
equilibrium density profiles of the different monomer species and  the incompressibility 
constraint allows us to infer the solvent density profile.
The net reduced monomer adsorption, $\Gamma^*$, is obtained as
$\Gamma^* = \int_0^\infty (n(z)-n_b)\sigma^3 dz$, where $n_b$ is the
bulk monomer density.  Integrals are solved on a grid (along $z$), up to some distance $H = 50\sigma$, beyond which
bulk conditions are assumed. Test calculations have verified convergence, i.e.,
that we have used a fine enough grid, and a large enough value of $H$.

Using this type of surface model, we have established that a prewetting transition 
can occur in a 300-mer polymer solution.
\begin{figure}[h!]
\begin{center}
       \includegraphics[width=7.8cm]{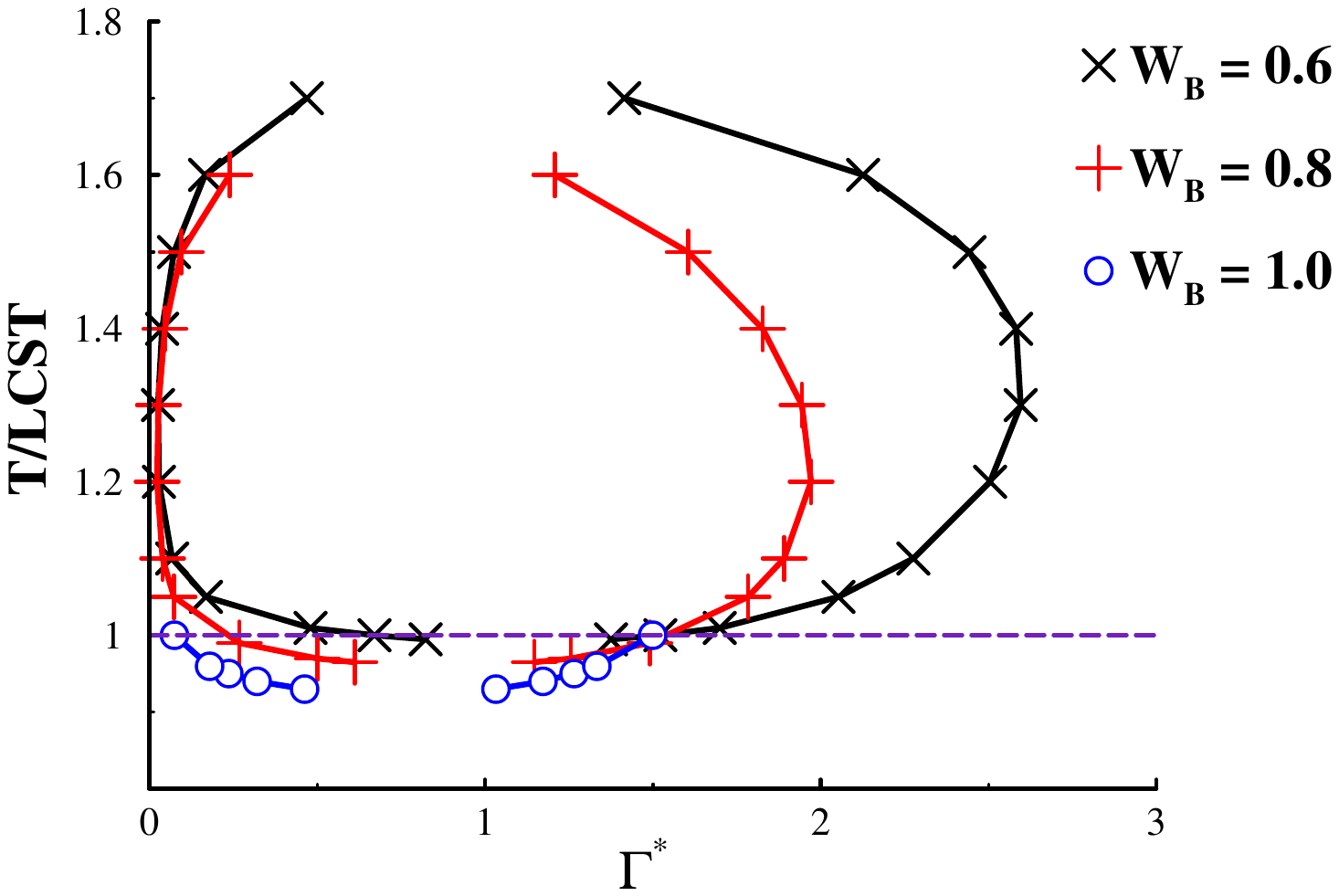}
       \caption{\footnotesize{Thin-thick coexistence curves, for a 300-mer polymer solution
           that displays an {\em LCST}. Crosses, plus signs, and circles denote
           various strengths of a short-ranged {\em B}-specific adsorption potential (see eq.(\ref{eq:wallpot})), with
           circles denoting the strongest attraction. Only the part at, and below, {\em LCST} (dashed line)
           is shown in the latter case, where the lower critical surface temperature
           is about 8 \% below the {\em LCST}. For PEO/water this would correspond to
           about 30 $^\circ$C.
}}
\label{fig:thinthick}
\end{center}
\end{figure}
In Figure \ref{fig:thinthick} we show the prewetting phase diagram, $\Gamma^*$
vs $T$, for different degrees of surface attraction, $W_B$. The temperature is scaled
with respect to the the bulk {\em LCST}.   The prewetting transition describes a closed loop
over the range, {\em UCST}$_{pw}$ to {\em LCST}$_{pw}$, which indicates that 
the prewetting line is completely detached from the bulk 2-phase line and there 
are no wetting temperature(s).   Furthermore, we see that both prewetting 
critical temperatures are below their bulk counterparts.  This confirms that 
prewetting transitions can occur in supracritical regions of the bulk fluid.
\begin{figure}[h!]
\begin{center}
       \includegraphics[width=7.8cm]{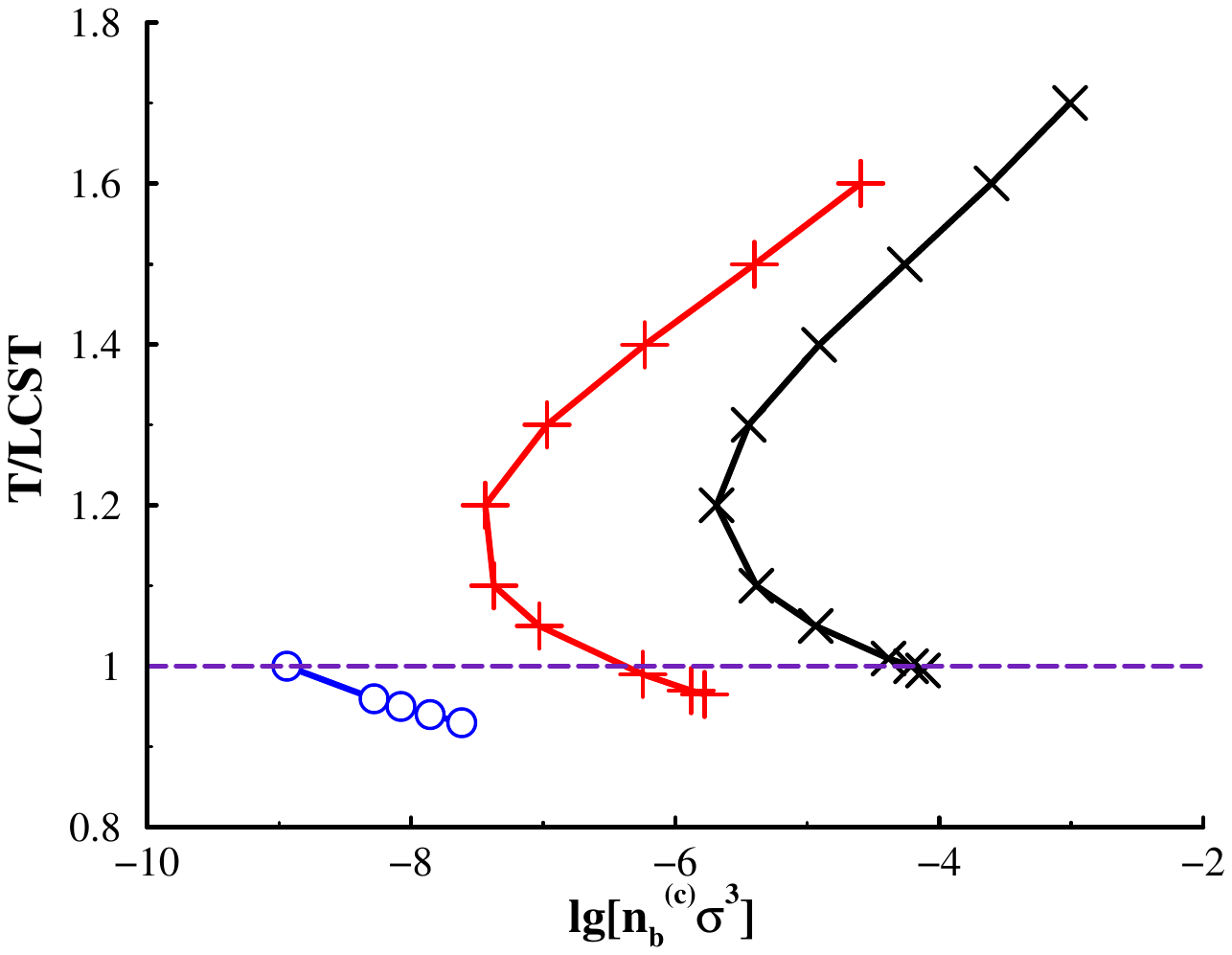}  
       \caption{\footnotesize {Prewetting line for 300-mer polymer solution
           that displays an {\em LCST}. $n_b^{(c)}$ denoted the bulk density at
           thin-thick coexistence. 
}}
\label{fig:prewettingline}     
\end{center}
\end{figure}
In  Figure \ref{fig:prewettingline} we show the prewetting line together with the dilute part of the
bulk coexistence region to more clearly illustrate the detached prewetting line.
It is not immediately obvious that one could not find a set of model parameters where 
the prewetting transition attaches to the bulk coexistence curve with two
(upper and lower) wetting temperatures, but we did not establish this scenario in 
the current study.  We also note that it is at least possible that, thin-thick transitions may take place at an inert 
surface without preferential attraction to any fluid species.
This is because of the unequal mutual interactions between species, i.e., the system
can minimise the wall-fluid interfacial tension by preferential adsorption of species
with weaker interactions.  For example, such a mechanism gives rise to CIPS 
in narrow inert pores. \cite{Xie:13,Haddadi:21b} 
We have so far only established thin-thick coexistence for surfaces that 
are somewhat more attractive to the solvophobic {\em B} monomers for the
current model. On the other hand, the surface
attractions required are short-ranged, and quite weak, so we do not rule out that
one might observe prewetting transitions for non-adsorbing surfaces with other interaction models.
\begin{figure}[h!]
\begin{center}
       \includegraphics[width=7.8cm]{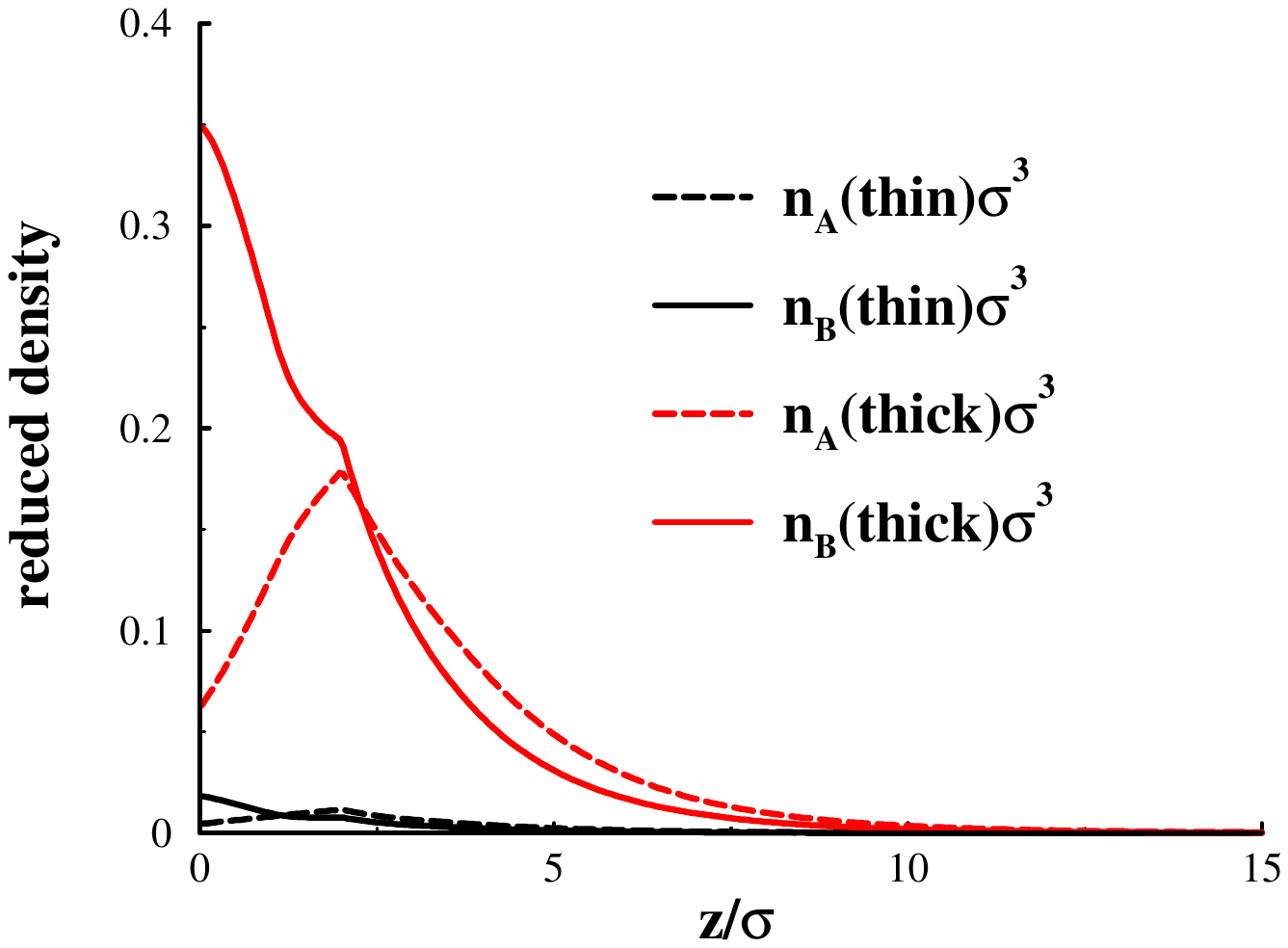}
       \caption{\footnotesize{Density profiles, for coexisting thin (black) and thick (red) phases,
           for a 300-mer solution, at the bulk {\em LCST}, with $W_B^* = 1.0$.
           Separate profiles for {\em A} (dashed) and {\em B} (solid) type
           are displayed.
}}
\label{fig:dprof}
\end{center}
\end{figure}
In Figure \ref{fig:dprof}, we give an example of density profiles for coexisting thin and thick phases; in this case
at a temperature corresponding to the bulk {\em LCST}.
These merge together at {\em LCST}$_{pw}$.

\begin{figure}[h]
\begin{center}
       \includegraphics[width=7.8cm]{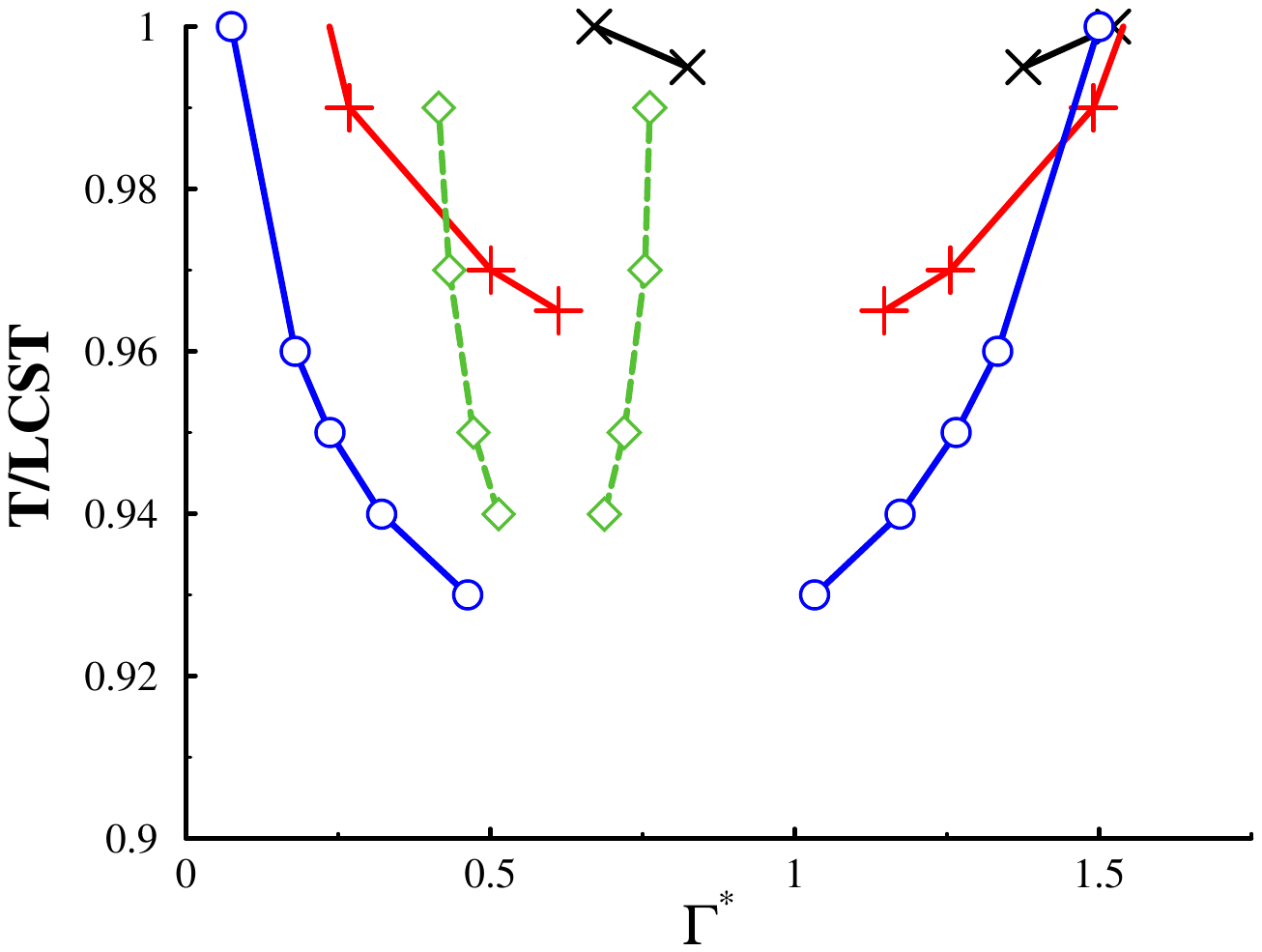}
       \caption{\footnotesize{Thin-thick coexistence curves, 300-mer polymer solutions, as
           well as a 100-mer polymer solution, belowe their respective bulk {\em LCST}.
           The notation for the 300-mer cases is the same as in Figure \ref{fig:thinthick}, whereas
           100-mer coexistence, with $W_B^* = 1.5$ is indicated by diamond symbols and dashed
           lines (the latter to guide the eye).
}}
\label{fig:rvar}
\end{center}
\end{figure}

The transitions will also depend upon the length of the polymer chains. In Figure \ref{fig:rvar}, we compare our results
for 300-mer mixtures, with a corresponding phase diagram for a 100-mer mixture.
Adsorption of polymers is a cooperative process, and with shorter polymers a stronger
surface affinity of {\em B}-type monomers is required, in order to
push the lower critical surface temperature below the bulk {\em LCST}.
In this case, we have set $W_B^* = 1.5$ for the 100-mers, which leads to
a similar $T/LCST$ ratio at the lower surface critical temperature, as
for 300-mers, with $W_B^* = 1.0$. We also note
a smaller overall difference between thin and thick phases, for shorter
polymers, throughout the demixing regime. 

\begin{figure}[h!]
\begin{center}
       \includegraphics[width=7.8cm]{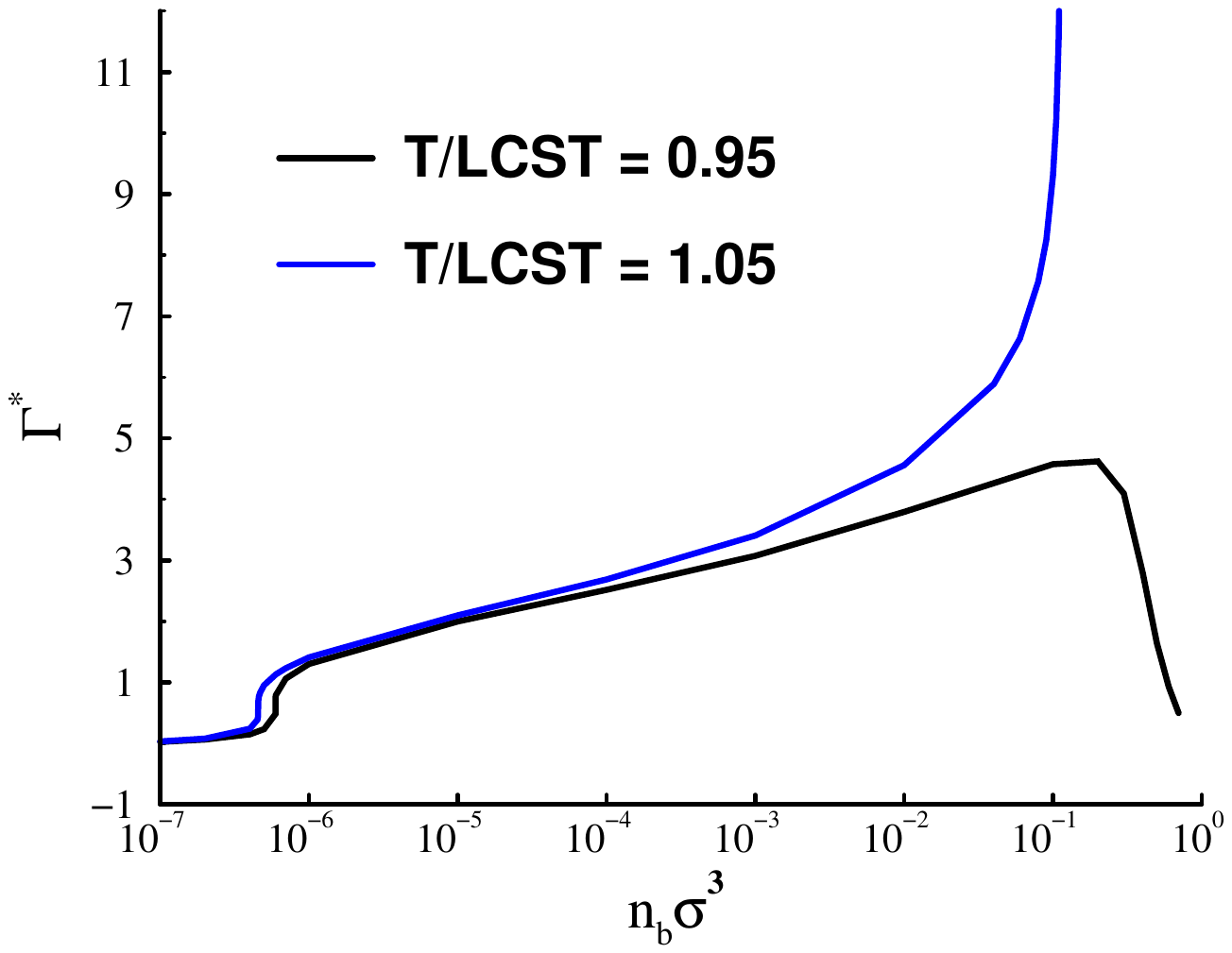}
       \caption{\footnotesize{The dependence of net adsorption on bulk concentration above and
           below the {\em LCST}. The ``jump'' at low concentrations signifies the first-order
           prewetting transition. The results are for a 100-mer solution, with $W_B^* = 1.5$.
}}
\label{fig:ads}
\end{center}
\end{figure}
Finally, in Figure \ref{fig:ads} we illustrate a fundamental difference between ``ordinary'' prewetting (above {\em LCST}) and
supracritical prewetting (below {\em LCST}). In the former case, the net adsorption diverges as we
increase the bulk phase concentration towards the bulk coexistence line.  For supracritical
adsorption, the thickness remains finite for all bulk concentrations and, indeed, the net adsorption
decreases at some point, as the bulk concentration increases, due to polymer depletion and the
lower energy per monomer in the bulk. 

This work thus confirms that first-order prewetting surface transitions can occur even in 
situations where the adjacent bulk fluid is outside its 2-phase region (supracritical).
In this case, it has been shown to occur for for mixtures that exhibit
a {\em LCST}.  However, our results imply that coexisting surface phases don't rely on the existence of a 
"corresponding" metastable bulk phase.  This latter type of description is 
often used in conjunction with the assumption of localised thermodynamics
in the inhomogeneous surface phase, which is clearly not justified when the spatial extent of the relevant fluid structures
are of the same order or smaller than the range of interactions.  One signature of this
{\em supracritical prewetting}  is that, as the bulk phase increases in concentration, there is no divergence in the
net adsorption of the thick phase, as occurs in complete wetting.  
Finally, it is of interest to note that all interactions in our model are isotropic.
In many ``real'' solution mixtures, there are also anisotropic interactions and, as
argued by Karlstr{\"o}m and co-workers\cite{Karlstrom:85,Andersson:85}, the
difference in solvophilicity between {\em A} and {\em B}  may be a result 
of a shift in polarity or hydrogen-bonding ability to the solute species.
These are orientation-dependent pair interactions.
The intrinsic degeneracies that distinguishes the {\em A} and {\em B}
may then be viewed as a coarse-grained description of
such differences. Moreover, our treatment here demonstrates that
there are no {\em formal} requirements of anisotropicity, in order
achieve the behaviours that we have investigated.

 J.F. acknowledges financial support by the Swedish Research Council.


\bibliography{poly}


\end{document}